\documentclass[10pt]{article}
\usepackage{wws2023} 
\usepackage{url}
\usepackage{latexsym}
\usepackage{amsmath, amsthm, amsfonts}
\usepackage{algorithm, algorithmic}  
\usepackage{graphicx}
\usepackage{lipsum} 
\usepackage{fancyhdr}
\usepackage[headheight=1in,margin=1in]{geometry}
\usepackage{hyperref}
\usepackage[LY1]{fontenc}
\usepackage[utf8]{inputenc}

\title{Measuring Americanization: A Global Quantitative Study of Interest in American Topics on Wikipedia}

\author{
  Piotr Konieczny \\
  Hanyang University,\\ South Korea\\\And
 Włodzimierz Lewoniewski \\
  Poznań University of Economics \\ and Business, Poland \\}

\begin{document}
\maketitle
\thispagestyle{fancy}

\begin{abstract}
We conducted a global comparative analysis of the coverage of American topics in different language versions of Wikipedia, using over 90 million Wikidata items and 40 million Wikipedia articles in 58 languages. Our study aimed to investigate whether Americanization is more or less dominant in different regions and cultures and to determine whether interest in American topics is universal.
\end{abstract}


\section*{Introduction}
Scientific publications on bias in Wikipedia have mainly focused on unequal contributions, such as gender and age bias, as well as the global digital divide. It is widely acknowledged that Wikipedia is biased toward Internet pop culture and that this bias reflects the systemic bias of global culture. The reasons for this bias can be explained by discourse and structural effects, with the latter being related to socioeconomic/demographic and cultural factors such as the number of speakers of a particular language, and the digital divide. While socioeconomic/demographic factors are better understood, explanations for cultural factors are sparse. However, it is evident that local Wikipedias tend to emphasize "local heroes" and create imbalances, although this can also be seen as a potential for representing multi-cultural diversity. Differences in cultural coverage are an unintentional result of the differing interests and experiences of editors. It is challenging to explain the size and quality of different Wikipedias just by socioeconomic factors, and cultural and geographical context influences communities' common interests and, in turn, gaps and unbalance in coverage of individual Wikipedia projects.

This paper aims to measure the extent to which American culture has influenced Wikipedia articles in different languages. Wikipedia has more than 100 different language versions, each created by volunteers from different regions. By comparing the amount of content related to the United States in different language versions of Wikipedia and the popularity of those pages, we can measure the degree of Americanization worldwide. This study will help us understand whether American culture dominates more or less in different regions and cultures.

\section*{Selection of Wikipedia articles}
When it comes to finding Wikipedia articles related to a particular topic, one approach is to use information from the categories that Wikipedia uses to group pages on similar subjects. However, in some languages, the category structure can be overly detailed and complex, making it difficult to analyze directly \cite{lewoniewski2019}. There are over 10 million categories spread across various language versions, which can be used to describe articles at different levels of abstraction \cite{lewoniewski2019}. 

Another way to locate articles on a specific topic is to draw on information from related items from other knowledge bases like Wikidata. Wikidata contains a range of statements that can be used to determine an object's connection to a particular topic \cite{lewoniewski2022}, in particular we can align such objects to particular culture, country, or language. For instance, a set of Wikidata properties can be employed for this purpose, including P17 (country), P19 (place of birth), P27 (country of citizenship), P276 (location), P495 (country of origin), and more. By utilizing these properties, it becomes possible to determine the relationship between an object and a specific country, culture, or language \cite{miquel2019}. Wikidata also gives the opportunity to find names of related Wikipedia articles in different languages. Therefore examples of objects associated with the United States are Wikipedia articles on Donald Trump (Wikidata item Q22686) or The New York Times (Q9684). Such interconnection allows finding Wikipedia articles on a specific topic or subjects with certain characteristics.  For example, it is possible to measure the ratio of women and non-binary-gendered Wikipedia biographies to total Wikipedia biographies based on Wikidata \cite{konieczny2018}.

\section*{Results}

We conducted an analysis of over 90 million Wikidata items by combining the Miquel-Ribé and Laniado model \cite{miquel2018,miquel2019}, which is based on Hecht's cultural contextualization theory, and Konieczny's use of the Inglehart-Welzel cultural clusters model \cite{konieczny2020}. By using the semantic connections of Wikidata to Wikipedia, we were able to determine which of the over 40 million articles in 58 different language versions were related to American topics.

For each considered language edition of Wikipedia the following parameters were measured:
\begin{itemize}
    \item PPCRW - People in a Primary Country Reading that Wikipedia (that language version), 
    \item VPC - Views from the Primary Country,
    \item US RAS - United States-related articles share,
    \item US RAVS - United States-related articles views share
\end{itemize}
Primary Country refers to the country with the largest number of Wikipedia readers using the considered language edition of the encyclopedia.
  
\begin{figure*}[ht]
\begin{center}
  \centerline{\includegraphics[width=\textwidth]{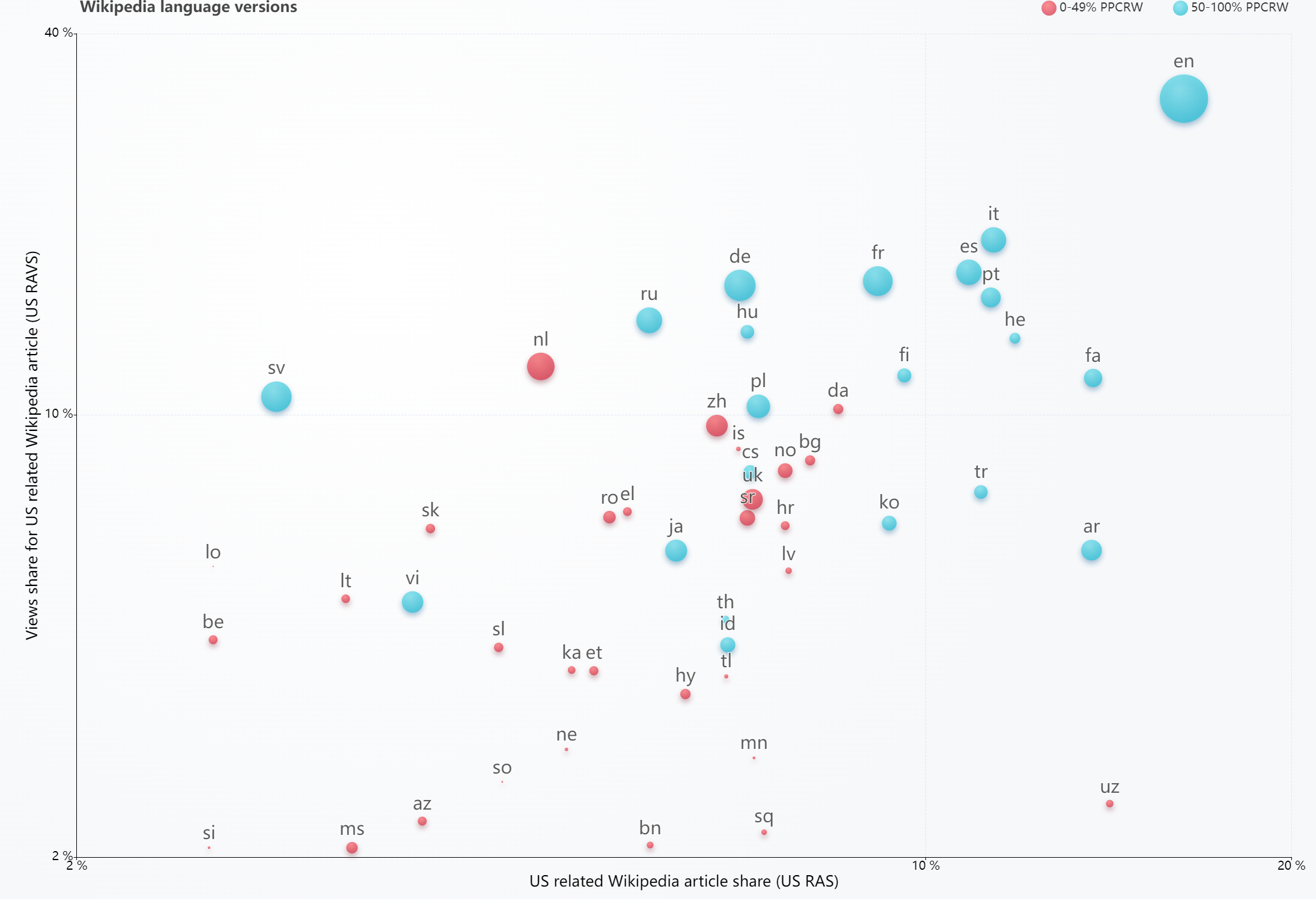}}
  \caption{Bubble chart with selected Wikipedia languages which uses US RAS and US RAVS values for location. The size of the item indicates the number of articles in Wikipedia language, color indicates the range of PPCRW values. More extended and interactive version on \url{https://data.lewoniewski.info/americanwiki}}
  \label{fig:teaser}
  \end{center}
\end{figure*}

The bubble chart depicted in figure \ref{fig:teaser} is a visualization where each language's position is determined by its US RAS and US RAVS parameters. The size of the bubble represents the number of articles contained in that particular Wikipedia language version, while the color range indicates the values of PPCRW. Languages with PPCRW values between 0-49\% are marked in red, while those with values ranging from 50-100\% are blue. To make the chart more readable and easier to understand, a logarithmic scale with limited values has been used. The chart shows that the languages marked in blue tend to have higher values of US RAS and US RAVS parameters when compared to those in red. Additionally, blue-colored Wikipedia editions generally have a larger number of articles compared to their red counterparts. 

Particular values of the considered parameters are presented in the table \ref{table1} for Wikipedia languages with at least 1 million articles.

Our findings indicate that Western, developed countries show greater interest in American topics, while the rest of the world appears less Americanized. Our data further suggest that the Americanization of Latin America is very significant, as at 16\% (popularity) and 11\% (article count) it is higher than the values for Europe. The latter have been aggregated, following the Inglehart-Welzel model, into several clusters: Catholic Europe (10\% popularity, 6\% article count), Protestant Europe (without English-speaking countries; 9\% and 6\%, respectively), and Orthodox Europe (7\% and 6\%, respectively). This conclusively confirms previous assumptions made in the literature on Americanization and related phenomena. This study provides the first global, quantitative confirmation of these issues, demonstrating that social science concepts once considered difficult to measure can be quantified through Wikipedia and Wikidata.

Furthermore, our research is relevant to discussions surrounding Wikipedia biases \cite[p.~77]{shaw2018,jemielniak2014},  as our findings shed light on how American topics are prioritized across different language versions of the platform. 

The findings of this study are most applicable to countries that are not multilingual and do not have significant diasporas. This is because the relationship between a country and its primary language is more straightforward in such cases. Additionally, the number of speakers of a language and the level of development of their country may influence the size and comprehensiveness of Wikipedia in that language. However, in developing countries, the digital divide can limit the representativeness of the results. Nevertheless, as non-English Wikipedias continue to grow and improve their coverage, the data should become more reliable over time.

\bibliographystyle{wws2023} 
\bibliography{paper}

\clearpage
\newpage

\begin{table*}[ht]
\centering
\begin{tabular}{|l|l|r|r|r|r|}
\hline
\multicolumn{1}{|c|}{\textbf{Abbr.}} & \multicolumn{1}{|c|}{\textbf{Language}} & \multicolumn{1}{c|}{\textbf{PPCRW}} & \multicolumn{1}{c|}{\textbf{VPC}} & \multicolumn{1}{c|}{\textbf{US RAS}} & \multicolumn{1}{c|}{\textbf{US RAVS}} \\ \hline
en        & English                          &  60\%        & 80\%      & 16.31\%                              & \textbf{31.58\%}                               \\ \hline
it            & Italian                      &  86\%        & 92\%      & 11.37\%                              & \textbf{18.89\%}                               \\ \hline
es           & Spanish                       &  70\%        & 60\%      & 10.85\%                             & \textbf{16.79\%}                               \\ \hline
fr         & French                         &  76\%        & 68\%      & 9.13\%                               & \textbf{16.26\%}                               \\ \hline
de        & German                          &  73\%        & 77\%      & 7.03\%                               & \textbf{16.01\%}                               \\ \hline

pt        & Portuguese                           &  54\%        & 80\%      & 11.31\%                              & \textbf{15.32\%}                              \\ \hline
ru         & Russian                         &  85\%        & 59\%      & 5.92\%                               & \textbf{14.10\%}                               \\ \hline
nl                 & Dutch                 &  34\%                                & 69\%      & 4.82\%                             &   \textbf{11.92\%}                              \\ \hline
sv         & Swedish                         &  57\%        & 89\%      & 2.92\%                               & \textbf{10.68\%}                               \\ \hline

pl       & Polish                           &  81\%        & 81\%      & 7.28\%                               & \textbf{10.31\%}                               \\ \hline
zh          & Chinese                        &  7\%                                 & 68\%      & 6.73\%                               & \textbf{9.61\%}                               \\ \hline

uk          & Ukrainian                        &  24\%                                & 85\%      & 7.20\%                              & \textbf{7.35\%}                                \\ \hline
ar                   & Arabic                & 60\%        & 70\%      & 13.69\%                              & \textbf{6.11\%}                               \\ \hline
vi         & Vietnamese                         &  71\%        & 92\%      & 3.78\%                               & \textbf{5.06\%}                               \\ \hline

\end{tabular}
\caption{Considered Wikipedia language versions with at least 1 million articles and their estimated parameters used in the study sorted by US RAVS (United States-related articles views share).}
\label{table1}
\end{table*}

\end{document}